# On Optimizing Resource Utilization in Distributed Connected Components

Mohsen Koohi Esfahani
0000-0002-7465-8003

***Abstract*—Connected Components (CC) is a core graph problem with numerous applications. This paper investigates accelerating distributed CC by optimizing memory and network bandwidth utilization.**

**We present two novel distributed CC algorithms, SiskinCC and RobinCC, which are built upon the Jayanti-Tarjan disjoint set union algorithm. To optimize memory utilization, SiskinCC and RobinCC are designed to facilitate efficient access to a shared array for all cores running in a machine. This allows execution of faster algorithms with larger memory bounds.**

**SiskinCC leverages the continuous inter-machine communication during the computation phase to reduce the final communication overhead and RobinCC leverages the structural properties of real-world graphs to optimize network bandwidth utilization.**

**Our evaluation against a distributed state-of-the-art CC algorithm, using real-world and synthetic graphs with up to 500 billion edges and 11.7 billion vertices, and on up to 2048 CPU cores, demonstrates that SiskinCC and RobinCC achieve geometric mean speedups of 29.1 and 16.8 times.**

## I. Introduction

Connected Components (CC) algorithm is a fundamental graph algorithm with various applications. Current state-of-the-art distributed CC algorithms have been primarily designed with the assumption that each CPU core serves as an independent processing elements (or "rank" in the Message-Passing Interface (MPI) paradigm). Although this approach simplifies programming, it may lead to inefficient utilization of memory space due to per-core private copy of the data structures. To overcome these limitations, we study the design of distributed-memory CC algorithms that optimize resource utilization, leveraging memory and network resources more efficiently. Our approach assigns ranks to machines, enabling the allocation of large-scale data structures that will be shared among all cores within a machine.

We build upon the Jayanti-Tarjan (JT) disjoint set union algorithm [1]. Unlike the Shiloach-Vishkin (SV) algorithm [2] and the Awerbuch-Shiloach (AS) algorithm [3], JT requires a single step of traversing of edges. This reduction in global synchronization steps makes JT well-suited for distributed processing. However, efficient execution of JT requires an array of size proportional to the graph rank (number of vertices) which becomes available in our distributed design.

We introduce two novel distributed CC algorithms based on the JT algorithm: SiskinCC and RobinCC. SiskinCC is based on the fact that each vertex is updated at most once in JT algorithm and leverages this feature to enable continuous communication between computation nodes. In this way, SiskinCC optimizes the performance by reducing the final communication overhead.

To optimize network bandwidth utilization, we introduce RobinCC that is based on the Root-Function JT algorithm, a generalized version of the JT algorithm. The Root-Function JT provides two key benefits: (1) arbitrary ordering between components, and (2) systematic selection of the root parent vertex in each component. Furthermore, RobinCC leverages the fact that a large portion of the vertices in real-world graphs are connected. RobinCC identifies the largest component and reduces the communication bandwidth for its vertices.

We evaluate SiskinCC and RobinCC in comparison to a state-of-the-art CC algorithm, for real-world and synthetic graphs of up to 500 billion edges and 11.7 billion vertices, and on up to 2048 CPU cores, 16 machines. Our evaluation reveals that SiskinCC and RobinCC achieve geometric mean speedups of 29.1 and 16.8 times, respectively, and reduce network bandwidth usage by a geometric mean of 31.2 and 36.8 times, respectively.

This paper is continued by reviewing the terminology and the related work in Section II. Section III presents the design of SiskinCC and RobinCC is introduced in Section IV. The implementation and analysis of the algorithms are discussed in Section V and the evaluation is presented in Section VI.

## II. Background & Related Work

**Terminology**. The immutable undirected static graph $G = (V, E)$ consists of a set of vertices $V$, and a set of edges $E$ between these vertices. Edges are unordered pairs of elements of $V$. The set of neighbors of vertex $v$ is denoted by $N_v$. The graph diameter is shown by $d$. CC algorithms utilize an array $L$, which maps a vertex to its component. Vertices $u$ and $v$ are in the same component, if and only if $L_u = L_v$.

**Related Work**. Major methods in CC algorithms are:

1. The *search-based CC* method [4] utilizes breadth-first search (BFS) or depth-first search (DFS) to identify the vertices in the same component. The computational complexity of this method is $\mathcal{O}(|V| + |E|)$.

2. The *label propagation CC* method [5–8] iteratively updates the vertex of a labels with the minimum label of its neighbors. The computational complexity of this method is $\mathcal{O}(d|E|)$. The *Shortcutting* technique [6, 7] allows each vertex to update the owner of its previous label with the new label, improving the complexity to $\mathcal{O}(|E|log|V|)$.

To accelerate propagation of the labels in skewed degree graphs, Thrifty [8] introduces the *Zero Planting* that assigns the zero label to the vertex with the maximum degree in the graph, enabling vertices in the largest component to quickly determine if further label update is possible, i.e., their label is greater than zero. To optimize the first iteration, Thrifty introduces the *Initial Push* technique that propagates the zero label from the max. degree vertex to its neighbors.

The Contour algorithm [13] determines the minimum value between endpoints of an edge and their parents and updates all of these values, facilitating faster convergence, i.e., reducing the number of iterations.

3. The *tree hooking based CC* and *disjoint set union based CC* algorithms [1–3, 9–11] initialize a forest of isolated vertices. The algorithm traverses the edges of the graph, and for each edge, the endpoint trees are merged if they belong to different trees. This method may require up to $\mathcal{O}(log|V|)$ passes over edges, resulting in a total computational complexity of $\mathcal{O}(|E|log|V|)$ in the CRCW model. These algorithms utilize an array to store the parent of vertices, where each vertex is initially its own parent. A parent is considered a root if its parent is itself. The parent array is updated during the execution to eventually create a tree for each component of the graph, connecting all vertices in that component.

The SV algorithm [2] consists of three steps in each iteration: (i) iterating over edges and hooking the first endpoint of each edge to its second endpoint if the parent of the second endpoint has not been updated and has a smaller ID, (ii) updating the parent of members of stars, and (iii) updating the parent of all vertices.

Bader et al. [10] present an optimized version of the SV algorithm which simplifies the process by removing the second step and updating the parent of each vertex to its root parent in the third step. LACC [12] and FastSV [11] present linear algebra based distributed algorithms based on Awerbuch-Shiloach [3] and SV, respectively.

Jayanti and Tarjan present a disjoint set union algorithm [1] that requires a single iteration over edges. For each edge, the algorithm performs the hooking of two endpoints by atomically finding their roots and hooking the one with the greater ID to the one with the lower ID. We explain JT in details in the following.

Afforest [14] divides the edges of each vertex into two sections which are processed in two iterations. Between these two passes, the algorithm identifies the largest component and skip processing half of its edges in the next pass. Alzi [15] suggests to improve Afforest by Shortcutting, Zero Planting, and Initial Push.

4. The *low diameter decomposition* method has also been used to determine CC [16, 17]. It works by iteratively identifying small diameter partitions [18] and contracting them, thereby reducing the number of vertices and edges. This method requires a computation complexity of $\mathcal{O}(|E|)$ in the CRCW model.

**JT CC Algorithm**. Algorithm 1 presents the JT-based CC

**Algorithm 1:** Jayanti-Tarjan (JT) [1] based CC

```
   Input: G(V, E), P[ ]
   /* Initialization                                   */
1  par_for v ∈ V
2  |  P_v = v;
   /* Hooking                                          */
3  par_for e(u, v) ∈ E
4  |  while true do
5  |  |  while P_u ≠ u do
6  |  |  |  u = P_u;
7  |  |  while P_v ≠ v do
8  |  |  |  v = P_v;
9  |  |  if u == v then
10 |  |  |  break;
11 |  |  if u < v then
12 |  |  |  if CAS(P_v, v, u) then
13 |  |  |  |  break;
14 |  |  else
15 |  |  |  if CAS(P_u, u, v) then
16 |  |  |  |  break;
   /* Pointer Jumping                                  */
17 par_for v ∈ V
18 |  while P_{P_v} ≠ P_v do
19 |  |  P_v = P_{P_v};
```

which utilizes the parent array $P$. A vertex $v$ is root, if $P_v == v$. $P$ is initialized by setting each vertex as its own parent.

The core concept involves traversing edges and merging the trees of the endpoints. This is accomplished by updating the parent of the tree root with the greater ID to the one with the lower ID. As the algorithm executes, the parent of a vertex may be updated *at most once*, indicating a new component with a lower ID. However, the parent of the vertex with the lowest ID in each component remains unchanged, and these vertices continue to act as roots. Each component is identified by the lowest ID of its vertices. The hooking step involves a single pass over all edges. For each edge, the algorithm identifies the root parent of both endpoints (Lines 5–8). If the root parents are the same, the endpoints are on the same tree. Otherwise, the root with the greater ID is merged to the the lower ID (Lines 11–16) using atomic Compare and Swap ($CAS$) operations.

The process of identifying the root parents and merging them is performed atomically. If another thread changes $P_u$ while the current thread has finished execution of Lines 5-6, the CAS operation on Line 15 fails and the $while$ loop needs to be executed again. It is similar for updating parent of $v$ based on the previously read $P_v$. However, if the parent value of the vertex with lower ID is updated by another thread while the parent value of the vertex with greater ID is being updated by the current thread, the correctness is not impacted.

Pointer jumping (Lines 17–20) is the last step and updates the parent of each vertex with its root parent. The two vertices $u$ and $v$ are considered connected if and only if $P_u == P_v$.

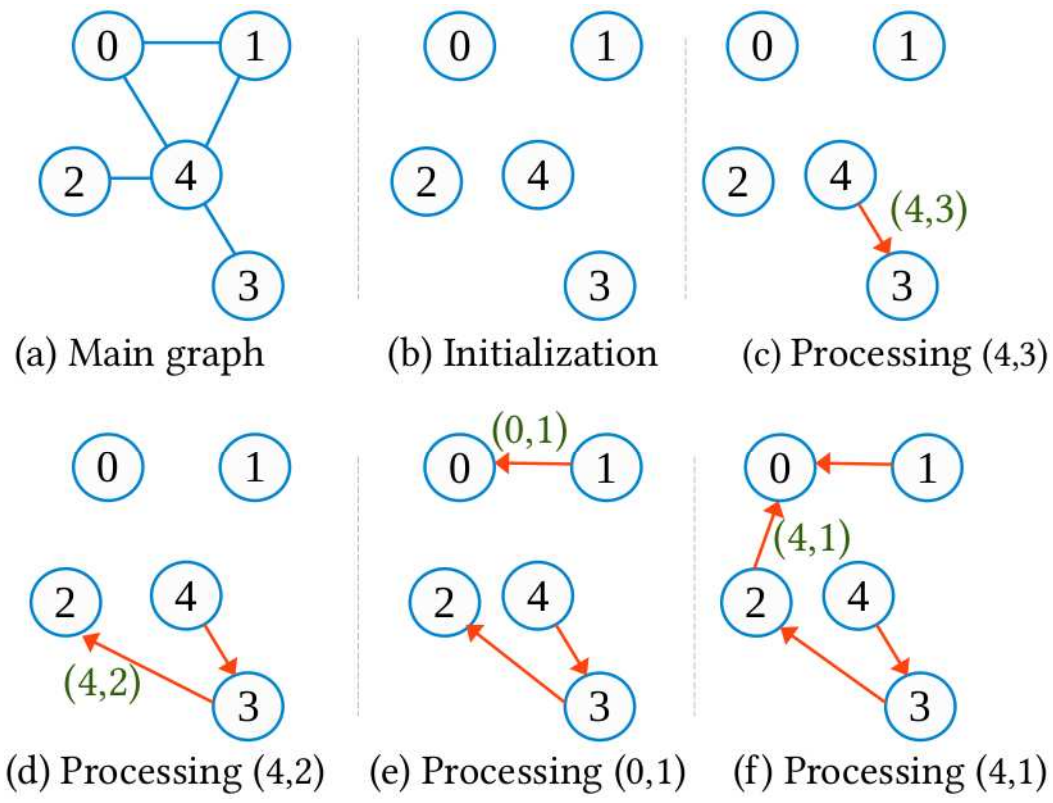

Fig. 1: An example graph and the changes of the parent array during the execution of JT algorithm. A red arrow line connects a vertex to its parent.

JT needs to process each edge in one direction, e.g, after processing $(u, v)$, processing $(v, u)$ has no impact. In other words, the input graph does not need to be symmetric, and for symmetric graphs, JT can skip processing half of the edges.

Figure 1 illustrates an example graph and the evolution of the parent array during execution of the JT algorithm. The red arrow lines indicate the parent of vertices. Self loops are not shown. Initially, each vertex is its parent (Figure 1.b). As the algorithm processes the edge (4,3), the lower endpoint (3) is set as the parent of the greater endpoint (4), as depicted in Figure 1.c. When the algorithm processes the edge (4,2), the root parents of endpoints 4 and 2 are 3 and 2, respectively, and 2 is set as the parent of 3, shown in Figure 1.d. The process continues for other edges as shown in Figure 1.e and f.

## III. SiskinCC

In comparison to other CC algorithms, JT has two major advantages: (i) a single pass over the edges, which translates to a single communication step between machines in a distributed setting, and (ii) removing the overhead of graph symmetrization. This led us to choose JT as the base connectivity method for SiskinCC and RobinCC, our distributed CC algorithms.

**High-Level Algorithm**. To utilize the JT (Algorithm 1) in a distributed setting, edges are divided between machines. Each machine updates its $P$ array while processing the assigned edges. A machine is designated as the *Reducer*, which is responsible for collecting data from other machines and aggregating them. Upon completion of processing edges by a machine, it traverses its $P$ array and sends pairs $(v, P_v)$ where $P_v \neq v$ to the Reducer. The Reducer processes the received pairs similarly to graph edges and applies a final pointer jumping step. The resulting $P$ array on the Reducer machine represents the final output.

In SiskinCC, we use a feature of JT algorithm to overlap compuatation and communication phases. In JT, the parent $(P)$ value of each vertex is updated at most once as the non-root vertices are never being updated. As an example, by processing edge (4,1) in Figure 1.f, the $P$ value of the endpoints in this edge are not updated, but $P_2$ is changed to 0. In other words, the first update to $P$ value of a vertex is the last one. We use this feature in SiskinCC and send vertex updates to the Reducer as soon as a $P$ value is changed. In this way, SiskinCC does not need to wait for completion of the computation phase to start the communication, minimizing the communication overhead and improving network bandwidth utilization during the computation phase.

**Algorithm 2:** SiskinCC distributed CC

```
   Input: G(V, E), P[ ], rank
   /* Partitioning the graph */
1  rank_parts = graph_partitioning(G, rank);
   /* Initializing P */
2  par_for v ∈ V
3  |  P_v = v;
   /* Processing edges and messages */
4  par_for tid ∈ threads
5  |  do
6  |  |  worklist = get_partition(rank_parts);
7  |  |  if rank.reducer ∧ worklist.empty() then
8  |  |  |  worklist = recv_msgs();
9  |  |  for (u, v) ∈ worklist do
10 |  |  |  while true do
11 |  |  |  |  while P_u ≠ u do
12 |  |  |  |  |  u = P_u;
13 |  |  |  |  while P_v ≠ v do
14 |  |  |  |  |  v = P_v;
15 |  |  |  |  if u == v then
16 |  |  |  |  |  break;
17 |  |  |  |  if u < v then
18 |  |  |  |  |  if CAS(P_v, v, u) then
19 |  |  |  |  |  |  send_msg(reducer, (v, u));
20 |  |  |  |  |  |  break;
21 |  |  |  |  else
22 |  |  |  |  |  if CAS(P_u, u, v) then
23 |  |  |  |  |  |  send_msg(reducer, (u, v));
24 |  |  |  |  |  |  break;
25 |  while (!conclude());
   /* Pointer Jumping */
26 if rank.reducer then
27 |  par_for v ∈ V
28 |  |  while P_{P_v} ≠ P_v do
29 |  |  |  P_v = P_{P_v};
```

**Pseudo Code**. Algorithm 2 shows the pseudo code of the SiskinCC. The graph is partitioned between machines based on the $rank$ values (Line 1). After initializing $P$ (Lines 2–3), each thread executes Lines 5–24 to process edges in the partitions (Line 6) and the received messages for the Reducer (Line 8). For each root vertex which is merged to another root with lower ID, a message is sent to the Reducer (Line 19 and 23). The computation on the Reducer

**Algorithm 3:** Root-Function JT CC

```
   Input: G(V, E), P[], f : V → Z≥0
   /* Initialization */
1  par_for v ∈ V
2      P_v = f_v;
   /* Hooking */
3  par_for e(u, v) ∈ E
4      while true do
5          while P_u ≠ f_u do
6              u = f^{-1}_{P_u};
7          while P_v ≠ f_v do
8              v = f^{-1}_{P_v};
9          if u == v then
10             break;
11         if f_u < f_v then
12             if CAS(P_v, f_v, f_u) then
13                 break;
14         else
15             if CAS(P_u, f_u, f_v) then
16                 break;
   /* Pointer Jumping */
17 par_for v ∈ V
18     u = v;
19     while P_u ≠ f_u do
20         u = f^{-1}_{P_u};
21     P_v = P_u;
```

machine is finished by a pointer jumping step (Lines 26–29).

## IV. RobinCC

To optimize network bandwidth utilization and reduce communication overhead of SiskinCC, in this section we present RobinCC which is based on the Root-Function JT CC, a generalized version of the JT algorithm.

**Root-Function JT CC**. In JT algorithm (Algorithm 1), the parent value of a root vertex $v$ is $v$. This prohibits (1) selecting the root vertex in a component and (2) applying ordering between components. To relax these limitations, we present Algorithm 3, the Root-Function JT CC algorithm, a generalized version of JT that utilizes a function $f$. Function $f$ maps a vertex ID to a non-negative integer number and specifies the root value for each vertex. As a result, the parent value of the root vertex $v$ is $f_v$.

To preserve the correct functionality of the algorithm it is required to ensure that each vertex is uniquely identified from its root value, i.e., $f$ must be an injective (one-to-one) function. We denote the inverse function of $f$ as $f^{-1}$. In Root-Function JT CC, for a vertex $v$, $P_v$ is initialized by $f_v$, and throughout the execution, this vertex is root if and only if $P_v == f_v$.

Algorithm 3 begins by initializing the $P$ array using the root function, $f$ (Lines 1–2). The hooking step involves processing all edges in the graph. For each edge $(u, v)$, the algorithm first identifies the root parent of the endpoints. This is done by iteratively applying the inverse function of $f$ to $P_u$ as long as $u$ is not a root vertex (Lines 5-6). Note that the input to $f$ is a vertex and input to $f^{-1}$ is a root value, i.e., a $P$ value. Similarly the root parent of $v$ is identified (Lines 7-8). If the root parent of endpoints are the same, no further action is taken for the current edge (Lines 9-10). Otherwise, the root parent with the greater $f$ value is merged to the one having the lower $f$ value (Lines 11-16).

The final step is pointer jumping, which involves identifying the root parent of each vertex using the $f^{-1}$ (Lines 18-21). Note that by deploying $f_v = v$, the Root-Function JT CC is transformed into the JT CC.

The Root-Function JT CC facilitates two major advantages:

- Root-Function JT allows selecting the root parent of a component, which is determined by the vertex with the lowest $f$ value, rather than the lowest ID as in JT.
- Root-Function JT also enables customizable ordering of components based on the lowest $f$ value of vertices in components, whereas in JT, components are ordered based on the lowest vertex ID they contain.

**High-Level Overview of RobinCC**. To optimize network bandwidth utilization, we design RobinCC by considering the characteristic structure of real-world graphs, which often exhibit a skewed degree distribution. This structure typically gives rise to a large component that encompasses a significant proportion of the edges and vertices [8, 19]. Notably, it has been observed that in real-world graphs, the largest component often contains more than 94% of the vertices [8].

In RobinCC, we use continuous communication only for the vertices in the largest component and instead of transferring pairs $(u, v)$, we send the ID of each new vertex has joined this component, resulting in half communication bandwidth for these vertices. To achieve this, it is required to recognize the largest component which is performed by applying Zero Planting [8] using the Root-Function JT (Algorithm 3).

RobinCC assigns a parent value of zero to the vertex with the largest degree. Consequently, the parent value of vertices that join the largest component become zero. Since zero is the smallest value between parent values with non-negative integers IDs, the root parent value of all vertices in the largest component (called *zero-converged vertices*) becomes zero. This facilitates recognition of these vertices and early transfer during the computation phase with half bandwidth usage. The root function ($R$) of RobinCC is defined as:

$$R_v = \begin{cases} 0 & v == \text{Max-Degree-VertexID}, \\ v+1 & \text{otherwise} \end{cases} \tag{1}$$

The inverse function of $R$, is specified by:

$$R_c^{-1} = \begin{cases} \text{Max-Degree-VertexID} & c == 0, \\ c-1 & \text{otherwise} \end{cases} \tag{2}$$

**RobinCC Pseudo Code**. Algorithm 4 demonstrates the RobinCC algorithm. Each machine has a different $rank$ value, which is used in Line 1 to create a binomial tree for efficient transfer of changed $P$ values. The $binomial_tree()$ function returns two values: (i) $com_parent$, the rank of the parent machine and (ii) $com_children$, an array of the children's ranks. Each machine sends messages to its parent and receives messages from its children. The machine with rank 0, called *root machine*, serves as the root of the binomial communication tree and does not transmit any messages. Additionally, the root machine functions will have the ultimate $P$ array upon completion of its execution. The machines that are placed as leaves in the binomial tree do not have any children and do not receive any message.

In Line 2, the input graph is partitioned between machines. We use an edge-balanced partitioning method, which assigns a block of continuous vertices to each partition while trying to keep the same number of edges in each partition, providing better load balance. Each machine is assigned a set of continuous partitions, which are processed by concurrent threads. The $owner(v)$ function is used to identify the rank of the machine who processes the vertex $v$.

The array $P$ is initialized in Lines 3-5 by identifying the vertex with maximum degree and the RobinCC's Root Function, $R$, Formula 1. In the $max_degree()$ function, each machine searches for the largest degree of vertices in its partitions. Then, the global max degree is identified by collecting the max degree of each machine. In Lines 6-8, the Initial Push technique is applied by traversing the neighbors of the max degree vertex and notifying their respective owners of the zero $P$ value for these neighbors. It is performed by calling $send_zero_converged()$ which allows other machines to identify the vertices with zero parent value during processing edges and benefit from their immediate transfer.

A machine send messages to its parent using two functions: (1) $send_zero_converged()$, which transfers a vertex whose $P$ value has recently been updated to zero, and (2) $send_parent()$, which sends a pair containing a vertex and its non-zero parent. The received messages are accumulated and accessed by two functions: (1) $recv_zero_converged_msgs()$, which returns an array of $(v, max_degree_vID)$ pairs, where v is a zero-converged vertex in one of the child machines, and (2) $recv_parent_msgs()$, which returns an array of $(u, v)$ pairs, indicating $P_u$ has been updated to $v$.

Lines 10–34 are executed in parallel by all threads. A $worklist$ array is retrieved (Lines 11-15), which contains the received pairs from the children or a partition of edges of the graph. All pairs in the worklist is processed similar to the Root-Function JT CC (Section IV), using the $R$ and $R^{-1}$ functions as defined in Formulas 1 and 2. The key difference is immediate sending of zero-converged vertices (Lines 26-27 and 31-32). The $conclude()$ function determines if the loop over worklists should be finished by checking if all child machines have finished and if all partitions assigned to this machine have been processed.

Pointer jumping and sending updated $P$ values to the parent machine is performed in Lines 35–43. By the end of this step, the $P$ array on the root machine contains the final result.

**Algorithm 4:** RobinCC distributed CC

```
Input: G(V, E), P[ ] , rank
   /* Communication network */
1  (com_par, com_children) = binomial_tree(rank);
   /* Partitioning the graph */
2  rank_parts = graph_partitioning(G, rank);
   /* Initializing P using R (Formula 1) */
3  max_degree_vID = max_degree(G);
4  par_for v ∈ V
5      P_v = R_v;
   /* Initial Push */
6  if rank == owner(max_degree_vID) then
7      par_for u ∈ N_max_degree_vID
8          send_zero_converged(owner(u), u);
   /* Processing edges and messages */
9  par_for tid ∈ threads
10     do
11         worklist = recv_zero_converged_msgs();
12         if worklist.empty() then
13             worklist = recv_parent_msgs();
14         if worklist.empty() then
15             worklist = get_partition(rank_parts);
16         for (u, v) ∈ worklist do
17             while true do
18                 while P_u ≠ R_u do
19                     u = R^-1_{P_u};
20                 while P_v ≠ R_v do
21                     v = R^-1_{P_v};
22                 if u == v then
23                     break;
24                 if R_u < R_v then
25                     if CAS(P_v, R_v, R_u) then
26                         if P_v == 0 then
27                             send_zero_converged(com_par, v);
28                         break;
29                 else
30                     if CAS(P_u, R_u, R_v) then
31                         if P_u == 0 then
32                             send_zero_converged(com_par, u);
33                         break;
34     while (!conclude(com_children, partitions));
   /* Pointer jumping & sending final values */
35 par_for v ∈ V
36     u = v;
37     while P_u ≠ R_u do
38         u = R^-1_{P_u};
39     if P_u == 0 ∧ P_v ≠ 0 then
40         send_zero_converged(com_par, v);
41     P_v = P_u;
42     if P_v ≠ R_v ∧ P_v ≠ 0 then
43         send_parent(com_par, v, u);
```

## V. Implementation & Analysis

**Implementation.** In SiskinCC and RobinCC, we utilize a background communication thread to handle message reception from the child machines and message transmission to the parent machine. The messages are organized in buffers, each containing up to 8 million vertex pairs. Each buffer is divided into 1024 blocks, enabling concurrent processing by threads.

For large-scale graphs, a vertex ID requires up to 8 bytes of memory space (Section VI). To optimize memory and cache utilization for accesses to the $P$ array, we employ a variable vertex ID size: 4 bytes for graphs with fewer than $2^{32}$ vertices, and 8 bytes for graphs with greater ranks.

To minimize communication overhead and maximize bandwidth efficiency, we use a compact vertex ID representation in communications, allocating $\lceil log_2(|V|)/8 \rceil$ bytes per vertex ID. Additionally, to accelerate communication between machines, we optimize memory allocation for asynchronous message reception from children, allowing senders to proceed without waiting for the target machine to allocate resources.

**Analysis**. The computational complexity of SiskinCC and RobinCC for processing edges remains similar to that of the JT algorithm, i.e., $\mathcal{O}(|E|log|V|)$ with a probability of at least $1 - 1/|V|$ [1]. The computation overhead for processing the received messages is $\mathcal{O}((m-1)|V|)$. Utilizing multiple machines, SiskinCC and RobinCC reduce the path from each vertex to its root parent is reduced.

Deploying $m$ machines and assuming that in the worst case, each machine updates the parent values of all vertices, the upper bound of the total communication volume is $\mathcal{O}((m-1)|V|)$. However, the total updates by all machines cannot exceed $|E|$ updates. Therefore, the communication volume is bounded by $\mathcal{O}(min(|E|, (m-1)|V|))$. This bound indicates that the communication volume in RobinCC, as well as SiskinCC, initially may increase with the number of machines, but remains constant when $m$ exceeds at most $|E|/|V|$, providing a good foundation for scalability. It may be notable that the binomial tree structure for communication in RobinCC does not change the communication volume, but it provides better concurrency and reduces communication time.

Each machine requires $\mathcal{O}(|V|)$ memory space to store the $P$ array. The functions $R$ and $R^{-1}$ are computed based on the input value and the max degree vertex ID and do not require storage in memory. Without considering the memory required for storing the graph topology, the total memory space required by RobinCC and SiskinCC on $m$ machines is $\mathcal{O}(m|V|)$.

**Zero Planting Impact**. The Zero Planting technique allows Thrifty to skip processing edges of the converged vertices in the rest of the current iteration and in subsequent iterations. In contrast, RobinCC needs to process all edges, as the neighbors of the zero-converged vertices must be merged with the zero component. As a result, RobinCC does not benefit from skipping edge processing, but it benefits from reduced communication bandwidth for converged vertices.

## VI. Evaluation

**Experimental Setup**. We conducted experiments on up to 16 machines, each equipped with two AMD 7702 CPUs, totaling 128 cores with a frequency of 2–3.4 GHz, 512 MB L3 caches, 1–2 TB of memory, and running CentOS 8. Machines are connected with dual Infiniband 100 Gb/s network interfaces to 7.2 Tb/s network switches.

TABLE I: Datasets

| Dataset | Type | $\mathbf{\|V\|}$ | $\mathbf{\|E\|}$ | $\mathbf{\|E_{sym}\|}$ | **\|CCs \|** |
|---|---|---|---|---|---|
| us-roads | Road | 23.9M | 57.7M | 57.7M | 1 |
| twitter-10 | Social | 41.7M | 1.5G | 2.4G | 1 |
| friendster | Social | 65.6M | 3.6G | 3.6G | 1 |
| clueweb09 | Web | 1.7G | 7.9G | 15.6G | 5,642,809 |
| g500 | Synthetic | 536.9M | 17.0G | 17.0G | 303,962,364 |
| msa10 | Bio | 1.8G | 25.2G | 41.7G | 628,505,933 |
| gitlab-all | VCH | 1.1G | 27.9G | 55.8G | 123,058 |
| eu-15 | Web | 1.1G | 91.8G | 161.1G | 1 |
| ms50 | Bio | 585.6M | 124.8G | 124.8G | 155,295,301 |
| wdc-12 | Web | 3.6G | 128.7G | 224.5G | 144,628,744 |
| swh-19 | VCH | 11.7G | 159.6G | - | 39,474,573 |
| msa200 | Bio | 1.8G | 500.4G | - | 221,467,156 |

The characteristics of the graph datasets used in our experiments are summarized in Table I, including the size of the symmetrized graph ( $|E_{sym}|$) and the number of connected components ($|CCs|$). Note that JT-based CC algorithms do not require the input graph to be symmetric (as discussed in Section II). Our datasets comprise a diverse range of graph types, including road networks, web graphs [20–26], social networks [27, 28], synthetic graphs [29], version control history graphs (VCH) [30, 31], and bio graphs [32].

We implemented SiskinCC and RobinCC in the `C` language utilizing `MPI`, `OpenMP`, and `libnuma`. Code is compiled with `gcc` 14.0.1 and `OpenMPI` 5.0.3 using the `-O3` optimization flag. The source code of SiskinCC and RobinCC is available online[1].

The input graph can be either in (i) WebGraph format [21] which is read using ParaGrapher library [33] or in (ii) CompBin format [33][2].

Graphs are stored in memory using the Compressed Sparse Row (CSR)which consists of an $offsets$ array with $|V|+1$ elements, each 8 bytes in size, and a $neighbors$ array with $|E|$ elements, each also of 8 bytes in size, as some datasets contain more than $2^{32}$ vertices.

We evaluate SiskinCC and RobinCC algorithms against FastSV [11] and Thrifty [8] state-of-the-art distributed and shared-memory CC algorithms.

**Performance, Scalability, and Network Efficiency**. Figure 2 compares the performance of FastSV, SiskinCC, and RobinCC for up to 2048 cores. Figure 3 compares the communication volume of the largest graphs in each category within our datasets. Note that FastSV requires the total number of processing elements (cores) to be a power of two, so we used the closest power of two for its experiments. Using up to 16 machines, FastSV could not load graphs larger than `msa10`. Also, some experiments for smaller graphs failed during the execution of FastSV. The execution time of FastSV does not include the symmetrization time.

As illustrated in Figure 2, SiskinCC achieves a speedup of 6.2–120.1 times compared to FastSV, with a geometric mean speedup of 29.1 times. The primary reason is single communication step which is overlapped by the computation. SiskinCC reduces the network bandwidth by 6.6–322.1 times compared to FastSV, with a geometric mean of 31.2 times. As discussed in Section II, other hooking-based CC algorithms require $\mathcal{O}(log|V|)$ passes over edges, while JT enables a single traversal of edges but with a work complexity of $\mathcal{O}(log|V|)$ per edge. This allows SiskinCC and RobinCC to strike a balance between reducing communication overhead and increasing computation cost in a distributed setting. Figure 3 shows up to more than two orders of magnitude reduction in communication volume for SiskinCC and RobinCC compared to FastSV.

[1]https://github.com/MohsenKoohi/SiPaR-SiskinCC-RobinCC
[2]https://github.com/MohsenKoohi/WG2CompBin

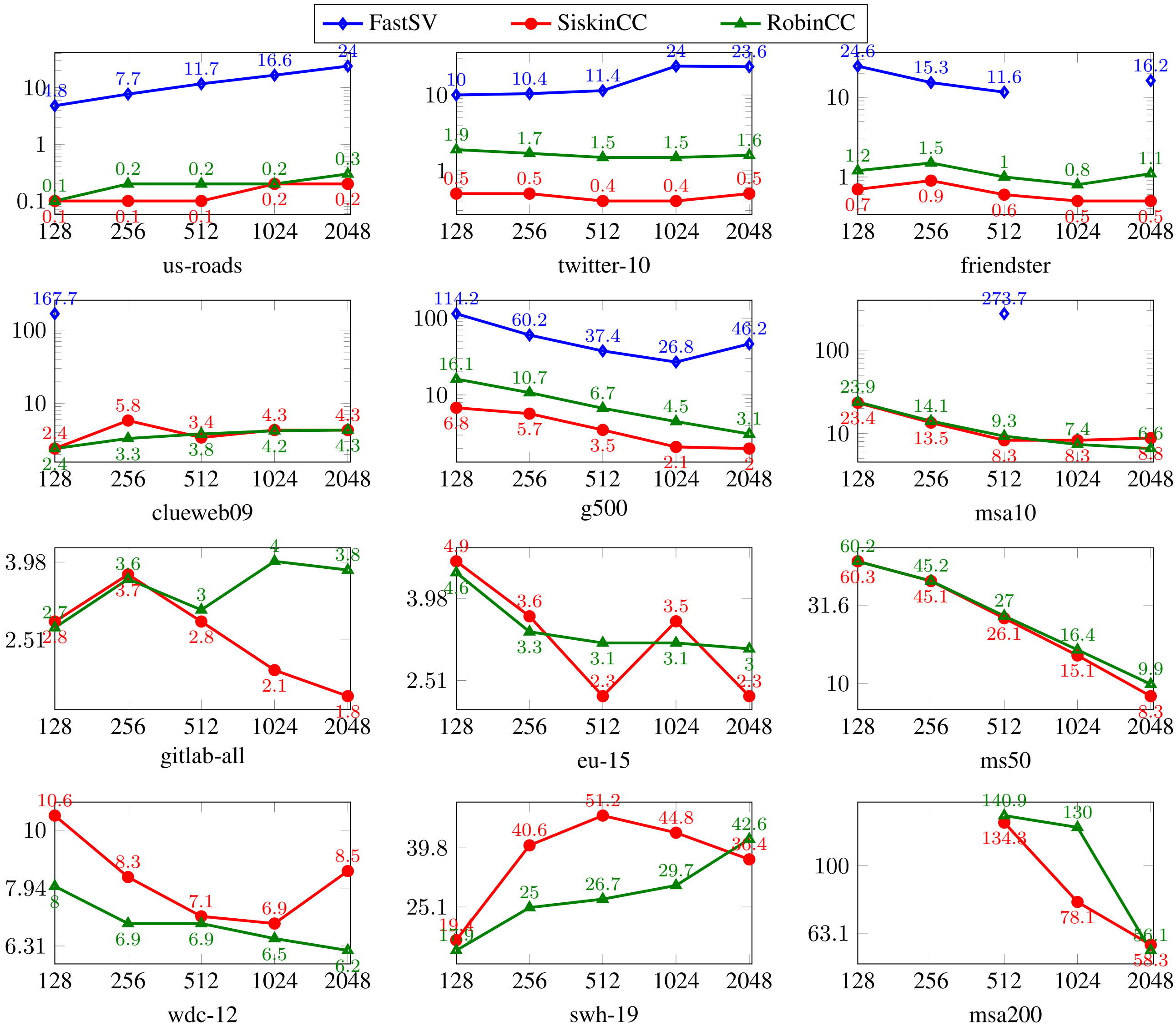

Fig. 2: Performance of FastSV, SiskinCC, and RobinCC. The x-axis indicates the number of cores and the y-axis shows the execution time in seconds.

RobinCC achieves 5.2–80.0 times speedup compared to FastSV, with a geometric mean speedup of 16.8 times. RobinCC reduces the network bandwidth by 12.0–241.5 times compared to FastSV, with a geometric mean of 36.8 times. For graphs smaller than `ms-50`, SiskinCC has usually a better performance than RobinCC, however, for large graphs such as `swh-19` and `wdc-12` RobinCC achieves a better performance. In general, for small graphs, the overhead of Zero Planting in RobinCC may not be compensated by provided faster communication. For the three largest graph RobinCC achieves a geometric mean speedup of 1.15 times over SiskinCC while reducing the network bandwidth by 1.32 times.

Figure 3 shows that RobinCC requires a smaller bandwidth than SiskinCC for almost all cases. It also shows that for `wdc-12` and `swh-19`, when running on 2048 cores, RobinCC uses a greater bandwidth than SiskinCC. This may be at-tributed to the high locality of these graph and the differences in communication mechanisms in the algorithms. In SiskinCC with a star communication topology, each machine directly sends its updated $P$ values to the Reducer. With a high locality level, the number of common vertices with updated $P$ values between machines is reduced, and these are sent directly to the Reducer. In contrast, RobinCC employs a binomial communication tree, where intermediate machines aggregate the received $P$ values and resend them. As the number of common vertices with updated $P$ values between machines is reduced, the replication of received messages into the sent messages is increased by RobinCC. However, this does not happen in direct communication between each machine and the Reducer in SiskinCC. The replication may also be increased by increasing the number of machines and increasing the depth of the communication tree. As a result, the binomial tree may increase bandwidth usage while providing concurrent communication and reducing communication time.

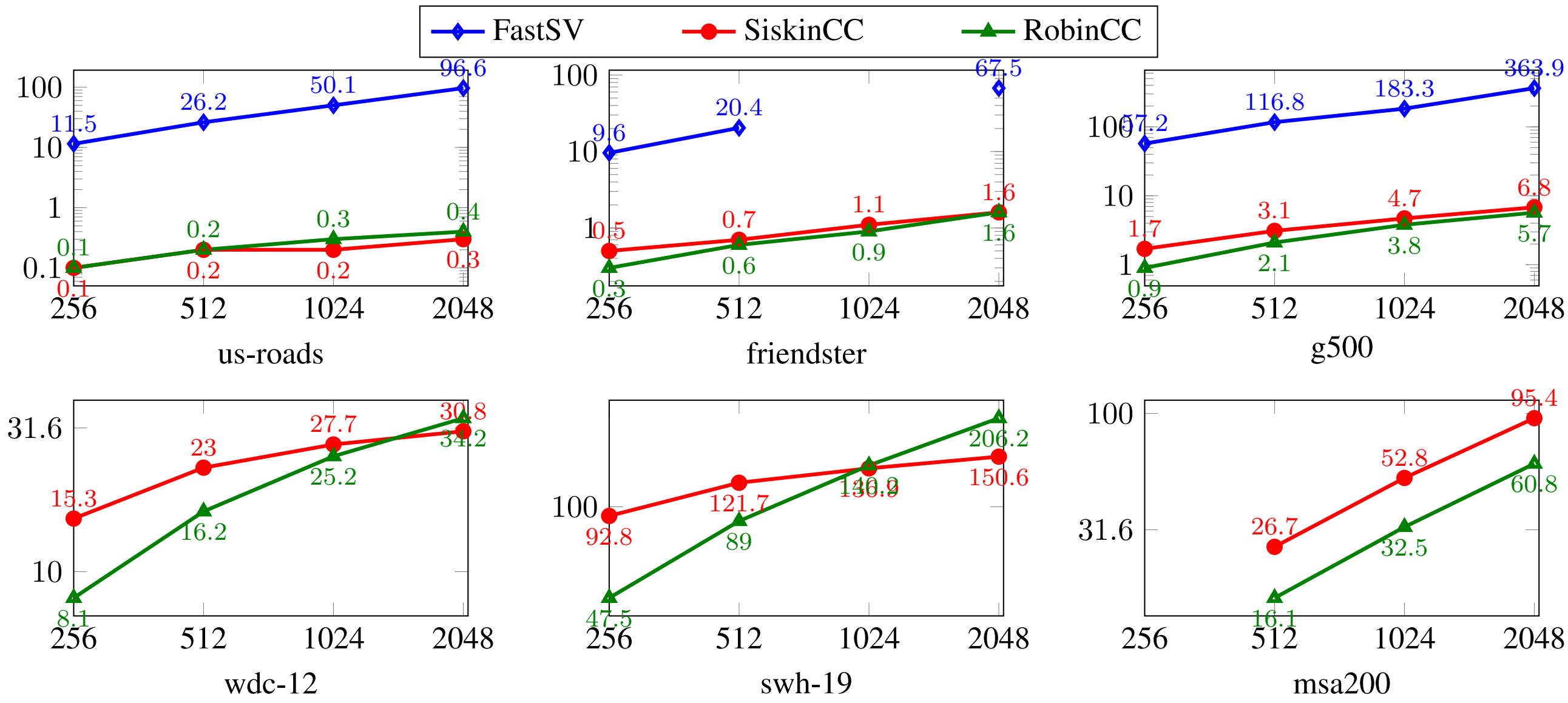

Fig. 3: Communication volume of FastSV, SiskinCC, and RobinCC. The line graph shows the used bandwidth in Gigabytes (y-axis) as a function of the number of cores (x-axis).

Considering the scalability of SiskinCC and RobinCC across all graphs in Figure 2, we observe two key patterns:

1. For graphs with high locality, such as web graphs and control version history graphs, the computation time is relatively short and the primary benefit of a greater number of machines is to accommodate the graph topology in memory. Since increasing the number of machines does not directly translate to a significant reduction in execution time, it is more efficient to use the smallest number of machines to reduce energy consumption and cluster utilization. This has been also observed in Contour [13].

2. In contrast, for graphs with low locality, such as bio graphs and synthetic graphs, the increased communication overhead is offset by the large size of computational workload, and the execution time is reduced through a good parallel efficiency. Furthermore, RobinCC has an upper communication bound of $\mathcal{O}(min(|E|, m|V|)$ (Section IV), which leads to reduction in communication and better scalability when the number of machines exceeds at most $|E|/|V|$.

**Memory Utilization**. Figure 4 compares the memory footprint of FastSV, SiskinCC, and RobinCC on four graphs and for running on 128, 256, and 512 cores. SiskinCC achieves 10.6–60.3 times reduction in memory consumption in comparison to FastSV, with a geometric mean of 20.1 times. RobinCC achieves 10.7–59.0 times reduction with a geometric mean of 20.0 times.

Considering the graph topology size in CSR format, FastSV requires 10–314 times of graph size as its memory space, with a geometric mean of 46.2 times. SiskinCC requires 1.6–7.1 times of the graph topology size, with a geometric mean of 2.5 times, and RobinCC requires 1.6–7.3 times, with a geometric mean of 2.6 times.

**Single-Machine Performance**. Table II presents a comparison of the single machine execution of RobinCC, SiskinCC, and FastSV with Thrifty [8]. Additionally, Table II shows the graph symmetrization time, a necessary preprocessing step for most CC algorithms, except those based on JT. Note that `us-roads`, `friendster`, `g500`, and `ms50` are symmetric and do not need this preprocessing time. The `msa200` graph exceeded the memory capacity of a single machine, making it impossible to load. Additionally, Thrifty's implementation had a limit of $2^{32}$ for the number of vertices, which prevented it from loading the `swh-19` graph. Comparing RobinCC and SiskinCC performance with Thrifty in Table II reveals a range of outcomes, from a 26.8× slowdown to a 48.2× speedup.

**Analysis of RobinCC's Execution**. Table III presents four timestamps in execution of RobinCC on 8 machines, R0–

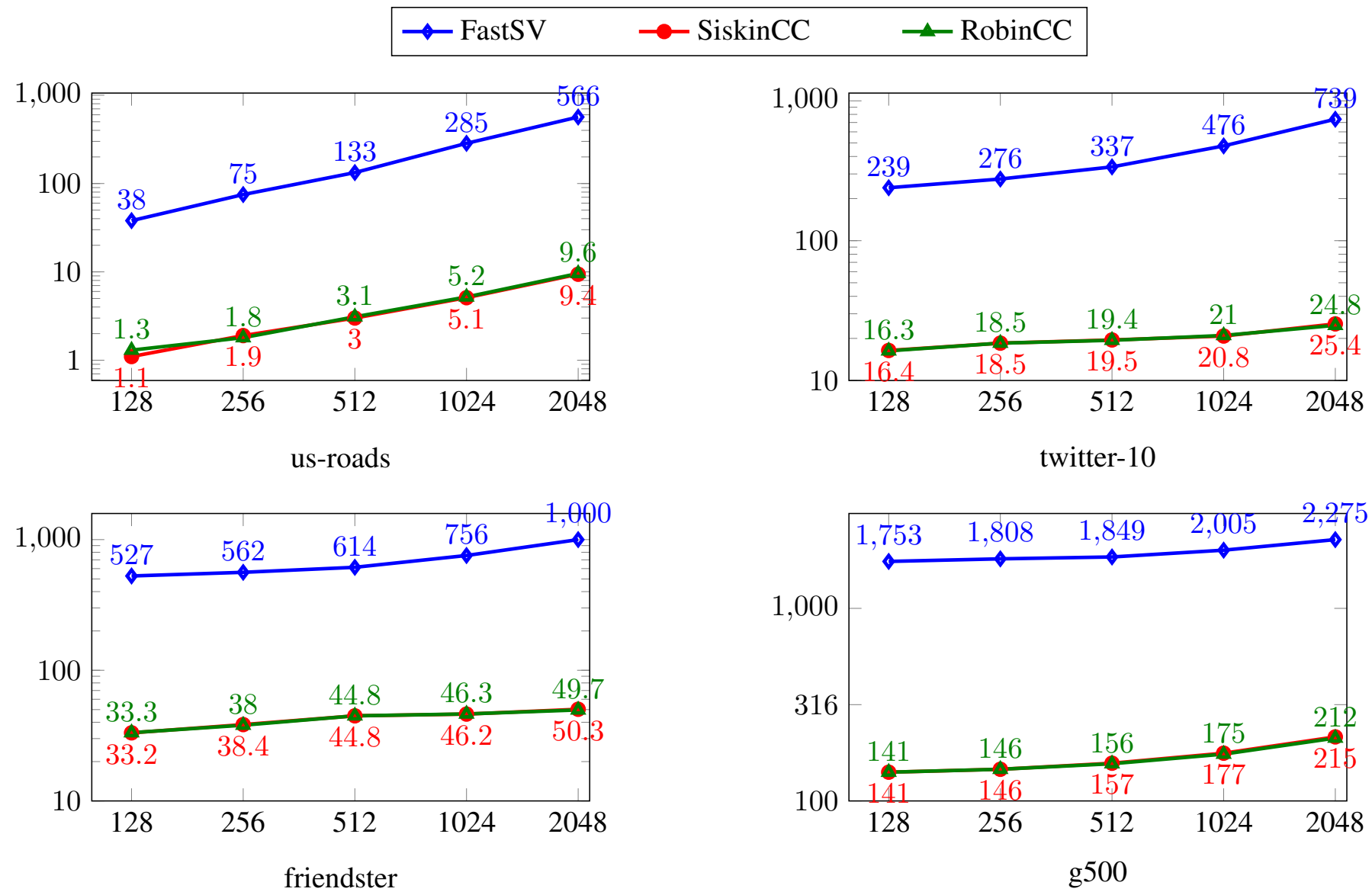

Fig. 4: Memory footprint of FastSV, SiskinCC, and RobinCC. The line graph shows the memory consumption in Gigabytes (shown on y-axis) as a function of the number of cores (shown on x-axis).

TABLE II: Single-machine execution. Numbers are in seconds. N: Not applicable. F: Failed loading.

| **Dataset** | **Symmetrize (s)** | | **CC Time (s)** | | | |
|---|---|---|---|---|---|---|
| | **Thrifty** | **FastSV** | **Thrifty** | **FastSV** | **SiskinCC** | **RobinCC** |
| **us-roads** | N | N | 4.8 | 4.8 | 0.1 | 0.1 |
| **twitter-10** | 12.0 | 64.6 | 0.1 | 10.0 | 0.5 | 1.9 |
| **friendster** | N | N | 0.1 | 24.6 | 0.7 | 1.2 |
| **clueweb09** | 30.0 | 290.3 | 2.5 | 167.7 | 2.4 | 2.4 |
| **g500** | N | N | 0.6 | 114.2 | 6.8 | 16.1 |
| **msa10** | 333.3 | F | 221.5 | F | 23.4 | 23.9 |
| **gitlab-all** | 25.1 | F | 1.5 | F | 2.8 | 2.7 |
| **eu-15** | 81.4 | F | 1.5 | F | 4.9 | 4.6 |
| **ms50** | N | N | 316.9 | F | 60.3 | 60.2 |
| **wdc-12** | 142.8 | F | 3.8 | F | 10.6 | 8.0 |
| **swh-19** | F | F | F | F | 19.4 | 17.9 |

TABLE III: Execution timestamps of RobinCC in seconds

| **wdc-12** | **R0** | **R1** | **R2** | **R3** | **R4** | **R5** | **R6** | **R7** |
|---|---|---|---|---|---|---|---|---|
| $T_1$. Edge Processing | 1.7 | 1.9 | 2.4 | 1.7 | 1.8 | 1.7 | 2.2 | 2.4 |
| $T_2$. Receive Messages | 6.4 | 1.9 | 2.6 | 1.7 | 5.7 | 1.7 | 3.6 | 2.4 |
| $T_3$. Pointer Jumping | 6.5 | 2.0 | 2.8 | 2.1 | 5.9 | 1.9 | 3.8 | 2.7 |
| $T_4$. Send Messages | 6.5 | 3.3 | 3.3 | 2.3 | 6.4 | 3.5 | 5.0 | 3.2 |
| **swh-19** | | | | | | | | |
| $T_1$. Edge Processing | 2.9 | 2.9 | 2.9 | 3.1 | 6.2 | 9.3 | 8.3 | 11.1 |
| $T_2$. Receive Messages | 29.2 | 2.9 | 7.1 | 3.1 | 24.9 | 9.3 | 16.0 | 11.1 |
| $T_3$. Pointer Jumping | 29.7 | 4.0 | 8.1 | 4.2 | 25.6 | 10.1 | 16.9 | 12.1 |
| $T_4$. Send Messages | 29.7 | 5.9 | 12.1 | 6.8 | 29.1 | 13.7 | 22.7 | 14.3 |

R7: *($T_1$) edge processing*: the time when processing of all assigned partitions is completed, *($T_2$) received messages*: the time of completion processing all incoming messages, *($T_3$) completion of pointer jumping*, and *($T_4$) completion of sending messages*. Note that processing edges and incoming messages are interleaved in RobinCC (Algorithm 4).

The binomial communication tree is shown in Figure 5, where R1, R3, R5, and R7 are leaves and do not receive any message. R0 is the root machine and does not send message.

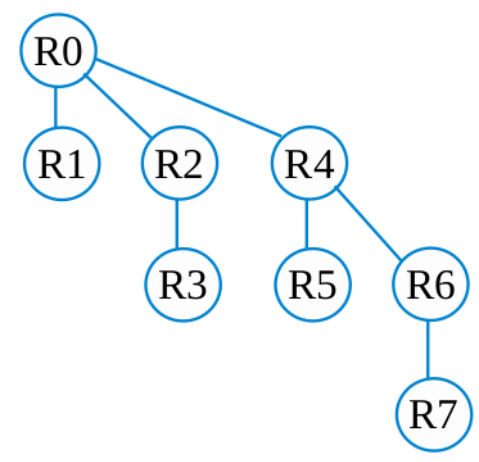

Fig. 5: Binomial communication tree

Table III shows that in processing `swh-19`, leaf machines spend a wider range of time (2.9–11.1 seconds) in processing edges ($T_1$), compared to 1.7–2.4 seconds in `wdc-12`. This may be attributed to the larger number of vertices in `swh-19` which degrades cache performance, but also offers potential for better partitioning schemes in `swh-19`. Additionally, `swh-19` has longer pointer jumping times ($T_3 - T_2$), due to its larger number of vertices.

## VII. Conclusion

This paper investigates methods for leveraging the memory space of modern server machines to design distributed Connected Components algorithms with optimized communication and performance. We introduce the SiskinCC and RobinCC distributed algorithms that enables threads within a machine to accelerate their computations by accessing a shared array. SiskinCC facilitates overlapping communication with computation and RobinCC optimizes network bandwidth utilization. Our evaluation results demonstrate that SiskinCC and RobinCC achieve speedups of up to 120.1 and 80.0 times, respectively.

**Future Work.** SiskinCC and RobinCC demonstrate the potential of designing more efficient distributed algorithms by optimizing resource utilization. This approach may be extended to other applications in high-performance and scientific computing, particularly those involving data-intensive algorithms that require significant resources and rely heavily on effective resource utilization for performance improvement.

We identify the following key areas as future work.

1. For certain graphs, such as `wdc-12` and `swh-19`, the binomial tree structure used for inter-machine communication can lead to increased communication time and volume, as discussed in Section VI. This suggests an opportunity to redesign the communication structure for better performance.

2. Although SiskinCC and RobinCC accelerates CC, a comparison of Figure 3 and Figure 2 reveals that a small portion of network bandwidth is utilized. Further investigations are required to identify methods for enhancing the overall performance by improving network bandwidth utilization.

3. RobinCC currently requires $\mathcal{O}(|V|)$ memory complexity on each machine (Section IV). Further research is needed to enable processing of graphs when the whole parent array ($P$) cannot fit in the memory of some machines, e.g., in heterogeneous environments. The goal is to host the largest possible portion of the $P$ array in memory and to share it between all threads within the machine.

4. The strategies used in designing SiskinCC and RobinCC are applicable to other asynchronous and deterministic graph algorithms that require a single pass over edges, such as Skipper Maximal Matching [34].

**Source Code**. The source code of SiskinCC and RobinCC is publicly available online on https://github.com/MohsenKoohi/SiPaR-SiskinCC-RobinCC.

## Acknowledgements

The computing resources of this work has been provided by NI-HPC (UKRI EPSRC grant EP/T022175/1). We used a non-training version of Llama 3.3 to enhance grammar and editing in this paper.